\documentclass[aps,twocolumn,superscriptaddress,showpacs,prl]{revtex4}
\usepackage{graphicx,amsfonts,amsmath,color,amsbsy,amssymb,subfigure}
\begin{document}
\title{Dynamic response of open cell dry foams}

\author{Buddhapriya Chakrabarti}
\email{buddhapriya.chakrabarti@durham.ac.uk}
\affiliation{Department of Mathematical Sciences, Durham University, Durham, 
DH1 3LE, United Kingdom.}

\author{Peter J. Hine}
\email{p.j.hine@leeds.ac.uk}
\affiliation{Soft Matter Physics Group, School of Physics and Astronomy,
University of Leeds, LS2 9JT, United Kingdom.}

\author{B. M. A. G. Piette}
\email{b.m.a.g.piette@durham.ac.uk}
\affiliation{Department of Mathematical Sciences, Durham University, Durham, 
DH1 3LE, United Kingdom.}

\date{\today}
\begin{abstract} 
We study the mechanical response of an open cell dry
foam subjected to periodic forcing using experiments and theory. Using the 
measurements of the static and dynamic stress-strain relationship, we derive
an over-damped model of the foam, as a set of infinitesimal non-linear springs, 
where the damping term depends on the local foam strain. We then analyse the 
properties of the foam when subjected to large amplitudes periodic stresses 
and determine the conditions for which the foam becomes optimally 
absorbing. 
\end{abstract}

\maketitle Foams of different types form an important part of our
daily lives having diverse applications depending on their material
properties, \textit{e.g.} density and
flexibility~\cite{Gibson:97,Wearie:99,Christensen:00}. Thus while open
cell polymeric low density foams are used for car cushions, semi-rigid
foams form packaging materials while rigid foams can be engineered to
provide thermal insulation. An important physical property of foams is
the efficiency with which energy is dissipated in these systems that
make them ideal as shock absorbers~\cite{Gibson:97,Ashby:00}. Since
material properties are intrinsically connected to the underlying
microstructure~\cite{Kraynik:08}, current research has focussed on its
classification~\cite{Kraynik:03,Kraynik:04} and an exploration of
mechanical properties using experiments~\cite{Papka:99,Gong:05},
numerical simulations~\cite{Kraynik:08,Kraynik:10} and
theory~\cite{Warren:88,Warren:91,Warren:97}.

X-ray microtomography~\cite{Gong:05} and scanning electron microscopy
of foams reveal their microstructure as connected random polyhedra
with their bulk mechanical properties governed by Gibbs elasticity of
the faces for closed cells, or nonlinear elasticity of struts for open
cell foams respectively~\cite{Warren:88}. Recently
Kraynik et al.~\cite{Kraynik:03,Kraynik:04} have performed detailed numerical
calculations to classify foam microstructure for monodisperse and
polydisperse foams. These simulations indicate that cells having an
average number of $\approx 14$ faces and $\approx 5$ edges are the
most probable polyhedra occurring in monodisperse foams in agreement
with experiments by Matzke~\cite{Matzke:46}. The surface energy of
such random foams have been estimated using scaling arguments and are
in good agreement with simulations~\cite{Kraynik:04}.

Analysis of strut deformation incorporating bending,
stretching~\cite{Gent:59,Warren:88}, and torsion~\cite{Warren:97} have
been employed to investigate the linear and nonlinear elastic behavior
of foams~\cite{Warren:91}. The nonlinearity arises as a result of the
coupling between geometry and strut elasticity. The micro-mechanics of
foams where constitutive models for struts are inherently nonlinear
have garnered recent interest~\cite{Kraynik:12}. The nonlinear
mechanical response has many fascinating features including strain
localisation~\cite{Gioia:01,Pampolini:08}, crushing~\cite{Papka:99,
Papka2:99}, energy dissipation, and shock mitigation~\cite{Avalle:01}.

\begin{figure}
\includegraphics[width=8cm]{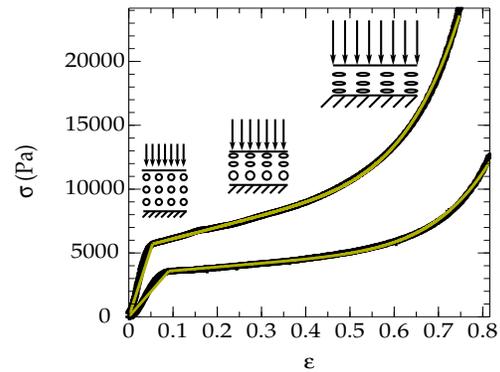}
\caption{Static compression data for foams, stress $Pa$ vs $\%$ strain. 
Two polynomial forms (see text) are fitted to the linear and non-linear 
part of the constitutive curve. Top curve, dense foam; bottom curve light foam.
}\label{Fig1}
\end{figure}

A complementary approach to the finite element simulations as a mean
to connect microscopic structure to mechanical properties is modelling
foams as a hyperelastic continuum~\cite{Ogden:72,Ogden:03}. Motivated
by these studies, \textit{in this Letter} we propose and validate a
phenomenological constitutive equation to model the mechanical
response of open cell solid foams subject to dynamic
loading~\cite{White:00}. While preparing this manuscript we came
across recent work in this area by Del Piero {\it et
al.}~\cite{DelPiero:09}. Our model is qualitatively different from
prior approaches in having a nonlinear constitutive equation and a
dynamic friction coefficient $\gamma(\epsilon)$ that is dependent upon
the local stress of the system $\epsilon$. Using this model we find
excellent agreement between experimental data for dynamic loading
experiments for two model foams, for low amplitude, low frequency
oscillations. This leads to several interesting results, namely the
identification of a scale factor $p$ that can be used to predict
system size dependence of foam properties within the validity of the
continuum approximation. We then explore, high amplitude, high
frequency forcing of model foams theoretically and obtain optimal
values of $p$ for energy dissipation and phase decoherence.


We carried out static and dynamic uniaxial compression tests for two
foams with densities $\rho_{d} \approx 41 kg/m^{3}$ and $\rho_{\ell}
\approx 19 kg/m^{3}$ (called dense and light respectively). 
The densities were measured using gravimetric
method with cuboid volumes measured using digital callipers and its
mass using a precision balance (Precisa $\alpha$B 320M) with an
accuracy of $1$ milligrams. Static compression tests were carried out
using an Instron model Extra between two steel plates having an upper
plate diameter $57.3$ mm, and a lower plate diameter $150$ mm. The
tests were carried out at $50 \%$ relative humidity and at a
temperature of $20^{\circ} C$. Cylindrical samples of light and dense
foams having diameter $d=37$ mm and $d=18.8$ mm and heights $L=21$ mm,
and $L=12.4$ mm respectively were used. A compressive displacement
rate of $0.167$ mm/s which translates to a compressive strain rate of
$7 \times 10^{-3} s^{-1}$. The voltage output was fed into a Pico
ADC-$20$ data logger and compressive stress-strain curves calculated.
 Dynamic
measurements were carried out in compression using a Rheometrics RSA
II solids analyzer. The samples were cut to a diameter of $19$ mm and
a length of $12$ mm and tested between two steel plates whose diameter
was $25$ mm. Initially a static compressive strain between $0.01$ and
$0.7$ was applied. The system was equilibrated for $1$ min following
which a dynamic test with a strain amplitude of $0.001$ and frequency
$f = 1$ Hz was applied. The storage and loss modulii $E^{\prime}$
and $E^{\prime \prime}$ as well as their relative phases
$\tan[\delta]$ were calculated for each value of base compressive
strain (increased in increments of $0.02$).

Fig.~\ref{Fig1} shows constitutive relation between the compressive stress 
$\sigma$ and strain $\epsilon$ for the static light (circles) and 
dense (diamonds) 
respectively. The experimental data is fitted (solid line) to a nonlinear 
function of the form
\begin{eqnarray}
\sigma &=&  \left\{
         \begin{array}{ll}
         \epsilon < \epsilon_0: & {\hat k_0}\, \epsilon \\
         \epsilon > \epsilon_0: & {\hat k_0}\, \epsilon_0 +
                        {\hat k_1}\, (\epsilon -\epsilon_0)
                       +{\hat K}\, (\epsilon -\epsilon_0)^6
\end{array}
         \right.\label{eq:staticstress}
\end{eqnarray} 
where, $\epsilon_0=0.053$, ${\hat k}_0=107896 Pa/m$,
${\hat k}_1=9162 Pa/m$, ${\hat K}=113919Pa/m$ for the dense foam
and $\epsilon_0=0.0905$, ${\hat k}_0=39570 Pa/m$, ${\hat k}_1=3164
Pa/m$ and ${\hat K}=54704 Pa/m$ for the light foam.
As is expected the stress-strain behavior is linear for
strains less than a critical threshold:
$\epsilon < \epsilon^{d}_0 \approx 0.054$ and
$\epsilon > \epsilon^{\ell}_0 \approx 0.09$.
For strains above the threshold the stress exhibits a large
plateau as the foam cells undergo ``collapse'', signalled by a large
change in strain at a nearly constant stress. Upon increasing the strain 
further the foam undergoes progressive densification as the opposite faces 
of foam cells meet, inducing the observed nonlinear stress-strain response.

We model the foam as a homogeneous material made of nonlinear elastic
elements obeying the force-extension relation given by
Eq.(\ref{eq:staticstress}), translationally invariant in $y$ and $z$
directions. The foam is held fixed at $x=0$ and a stress of the form 
\begin{equation}
F(t) = F_0 + F_A\, cos[\nu t] 
\label{eq:forcing}
\end{equation} is applied at $x=L$. 

\begin{figure}
\includegraphics[width=8cm]{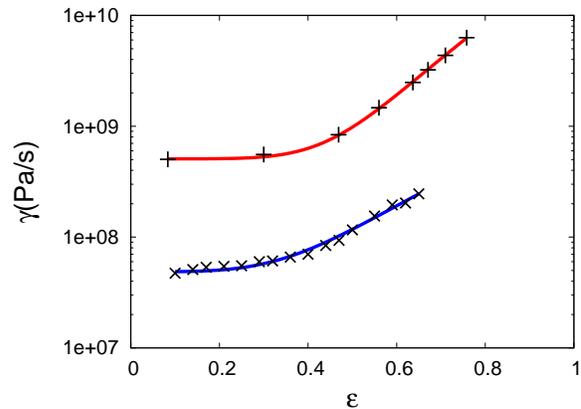}
\caption{Strain dependence of $\gamma$, dense foam ($+$) top curve and light foam 
($\times$) bottom curve respectively. Solid lines are theoretical fits to the data.}
\label{Fig2}
\end{figure}

As shown in the supplementary material, the equation describing the non-linear 
foams considered here is given by
\begin{equation}
\gamma \frac{du}{dt} =
\left\{ \begin{array}{ll}
         u_x > -\epsilon_0: & k_0\,u_{xx}\\
         u_x < -\epsilon_0: & k_1\,u_{xx}
            + K\,6\,(-\epsilon_0-u_x)^{5}u_{xx} \\
         \end{array}
        \right.
\label{eq_u_adim}
\end{equation} 
where $u(x)$ is the local displacement relative to the 
unstrained configuration of the
foam; $\gamma$ is a friction coefficient which, a priori, can
depend on the strain. We neglect the inertial term as in our set up it is 
$13$ orders of magnitudes smaller than the damping term
due to the low density of the foam.
Notice that the boundary conditions above are
such that $u(0)=0$, $u(x) < 0$ for compression  and $\epsilon=
-u(L)/L$. To determine $\gamma$, we have measured $\tan(\delta)$ for
the two foams using a dynamic strain of $0.001$ and
simulated these experiments using eq (\ref{eq_u_adim}). 
The strain dependent damping coefficient $\gamma$ was fitted to the 
experimental data using
\begin{eqnarray}
\gamma=\gamma_0 (1+ \gamma_1 \epsilon^{n})
\end{eqnarray}
with $\gamma_0=5.31981\,10^8 Pa/s$ $\gamma_1=60$ and $n=6$
for the dense foam and
$\gamma_0=4.87\, 10^{7}Pa/s$ $\gamma_1=22.62$ and $n=4$
for the light foam, with very good agreement as shown in figure \ref{Fig2}.

To study the properties of a foam subjected to a large amplitude periodic strain 
of period $T$, it is worth noting that eq (\ref{eq_u_adim}) can be rescaled as 
follow:
\begin{equation}
(1+\gamma_6\,u_y^6) \frac{du}{d\tau} =
\left\{ \begin{array}{ll}
       u_y > -\epsilon_0: & k_0^* \,u_{yy}\\
       u_y < -\epsilon_0: & k_1^* \,u_{yy} \\
 & + K^* \,6\,(-\epsilon_0-u_y)^{5}u_{yy}
         \end{array}
        \right.
\end{equation}
where, defining $p=T/\gamma_0 L^2$, $y = x/L$ and $\tau = t/T$ we have 
$k_0^* = k_0\,p$, $k_1^* = k_1\,p$, $K^* = K\,p$ and $F^*= F\,p$.
The scaling parameter $p$ allows us to relate the length scale, time scale and
damping parameter $\gamma_0$ of the foam. We thus see immediately that 
increasing the period of the load, $T$, decreasing 
the sample length $L$ or the foam friction $\gamma$ are equivalent in this 
rescaled system and correspond to a variation of $p$.
For the dense foam we have $k_0^*\approx1.382$, $k_1^*\approx0.1173$, 
$K^*\approx 1.459$ and $F^* \approx 1.28075\, 10^{-5}\, F/Pa$\

The amount of energy transferred, $E_{tr}$, to the foam by the excitation 
can be easily evaluated as
\begin{eqnarray}
\hat{E_{tr}} = \int_0^T F \frac{du}{dt} dt = 
            \frac{L}{p}\int_0^T F \frac{d{\hat u}}{d\tau} d\tau .
\end{eqnarray}
The energy dissipated by the foam, $E_{dis}$, on the other hand, is given by
\begin{eqnarray}
\hat{E_{dis}} = \int_0^L\int_0^T\gamma \left(\frac{du}{dt}\right)^2 dt dx = 
\frac{L}{p} \int_0^L\int_0^T\left(\frac{d\hat{u}}{d\tau}\right)^2  d\tau dy.\\
\end{eqnarray}
Then, the absorption quality of the foam will be given by the ratio 
$E_{dis}/E_{tr}$ which corresponds to the relative amount of absorbed energy.

To analyse the properties of a foam under large amplitude periodic stress, we 
have considered the dense foam subjected to a sinusoidal pulse 
$F(\tau)= 0.128*(1+\cos(\tau))$. The maximum stress $F^*=0.256$ correspond to 
a load of $20 kPa$ for the dense foam and as can be seen from fig 
\ref{Fig1}, this correspond to a strain $\epsilon=0.6$. We are thus 
probing the full range of the foam strain.



\begin{figure}
\includegraphics[width=10cm]{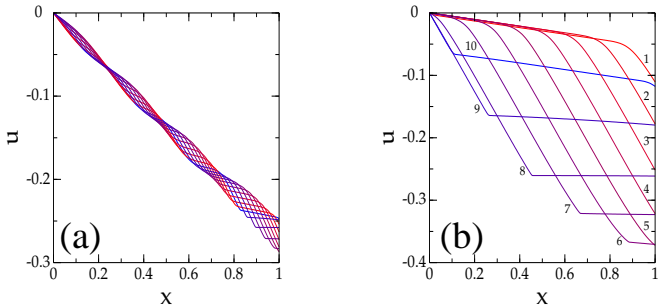}
\caption{Displacement $u(x)$ of a foam subjected to periodic excitation 
$F=0.128075*(1+\cos(t/T))$  during 1 period of oscillation. 
$k_0^*\approx1.382$, 
$k_1^*\approx0.1173$, $K^*\approx 1.459$ and $\gamma_1=60$. a) $p=0.01$, 
b) $p=0.05$. The sequence for the profiles is indexed in figure b.}
\label{Fig3}
\end{figure}

On figure \ref{Fig3} we present snapshots of the foam displacement at
regular interval during one period of oscillation, when subjected to
the sinusoidal pulse described above. Fig \ref{Fig3}.a corresponds to
$p=0.01$ while \ref{Fig3}.b corresponds to $p=0.05$. It is interesting
to notice that the displacement profile is not sinusoidal as one might
expect but instead progresses inside the foam as small plateaux. Small
values of $p$ correspond to large values of $\gamma$. One thus sees
that in that case, the amplitude of excitations are decreasing as one
moves towards the origin. When $p=0.05$, the foam deformation
progresses as a pulse, exhibiting large regions where the strain,
$du/dx$, is very small. At the base of the foam, one notices that the
strain is very small for the first half of the cycle and then large
and constant for the second half.

\begin{figure}
\includegraphics[width=8cm]{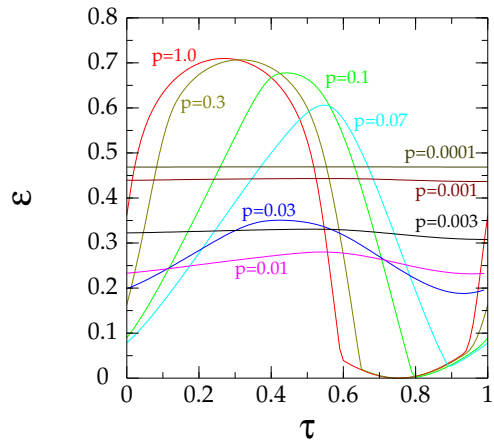}
\caption{Strain ($\epsilon=-u(L)/L$) of a foam
subjected to periodic excitation
$F=0.128075*(1+\cos(t/T))$  during 1 period of oscillation.
$k_0^*\approx1.382$, $k_1^*\approx0.1173$, $K^*\approx 1.459$ and
$\gamma_1=60$.
$p=1.0$, $0.3$, $0.1$, $0.07$, $0.03$, $0.01$, $0.003$, $0.001$, $0.0001$.}
\label{Fig4}
\end{figure}

\begin{figure}
\includegraphics[width=8cm]{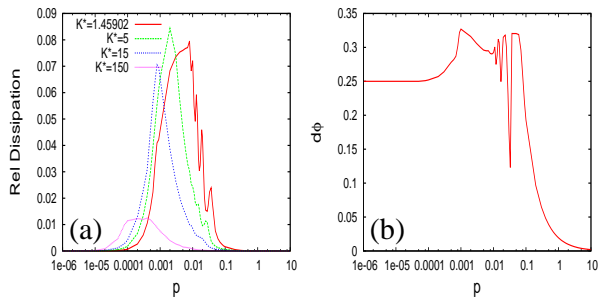}
\caption{Dependence on $p$ of the relative energy absorption (a) and  Phase
difference (b) of a foam  subjected to periodic excitation
$F=0.128075*(1+\cos(t/T))$.
$k_0^*\approx1.382$, $k_1^*\approx0.1173$ and $\gamma_1=60$. 
$K^*\approx 1.459$ for (b)}
\label{Fig5}
\end{figure}

On figure \ref{Fig4} we present the time evolution of the foam strain
during one cycle. For large values of $p$, the strain oscillates
between $0$ and $0.7$ while for very small value it is nearly
constant, just below $0.4$. In the former case, the foam does not
dissipate much and the load is transferred nearly unaffected by the
foam. In the later case, on the other hand, the foam is very
dissipative and the foam adopts a constant strain effectively
averaging over the load it is subjected to. It is also interesting to
note that there is a range of values around $p=0.01$ where the strain
of the foam is nearly constant but with a very small value ($\approx
0.25$). This is the value of $p$ such that the inside deformation of
the foam are attenuated over a length scale comparable to the
thickness of the foam. As a result, the foam is optimally
absorbing. This is well illustrated on figure \ref{Fig5} where we
show the relative absorption of the foam as a function of $p$ for
various values of the nonlinear elasticity parameter $K^*$. The
maximum, which does not vary much with $K^*$, lies mostly in the range
$p = 0.01$ to $p = 0.001$. When $K^*=150$ the foam is very stiff 
and the penetration much smaller, explaining the small absorption.
Moreover, we can see from Figure
\ref{Fig5}.b that the phase difference between the excitation and the
foam deformation is larger than $90$ degrees in the region of maximum
absorption. There is thus a partial phase opposition between the
stress and the induced strain which results in the nearly constant low
strain of the foam as described above.

We have also considered a square pulse given by $F(\tau)= 0.256 (0 \le
\tau < t_{PW})$ and $F(\tau)= 0.128 (t_{PW} \le t < 1)$ and we have
observed a similar behaviour of the foam. While of the oscillation
profile inside the foam changed, the absorption properties of the foam
was very similar to the one obtained for a sinusoidal excitation.

In conclusion, we have derived an heuristic
continuum model of dry foam deformation characterised by a non linear
elastic response as well as a strain dependant friction term. We have
fitted the value of our model to the experimental data of two foams
which both exhibited a 6th order term for the inelastic term but a 6th
and 4th power term for the strain dependence of the friction
coefficient.

We have also studied the properties of the foam under large amplitude,
periodic oscillation and have characterised a regime where the foam is
optimally absorbing. The optimal range is frequency dependant and can
be achieved by changing the thickness or the friction of the
foam. This is potentially relevant for the design of protective foam
pads to protect fragile objects during transport. One should use a
foam where the amplitude of the stress is such that it strains the
foam in its non-linear regime. This will depend mostly on $k_0$ and
$K$. If one can estimate the frequency of the excitations to which the
object to protect will be subjected, for example $1Hz$ for a parcel
being transported, then one can determine the optimal thickness of the
protective pad after measuring $\gamma$ for the selected foam.

\section{Acknowledgements}
\label{acknowledgements}

BC thanks EPSRC for support via grant EP/I013377/1 and BBSRC via grant BB/J017787/1.

\end{document}